\documentclass[aps,pra,twocolumn,groupedaddress,showpacs,nofootinbib,longbibliography,notitlepage]{revtex4-2}
\usepackage{etex}
\usepackage{amsmath,amssymb,amsthm}
\usepackage[colorlinks=true,citecolor=blue,urlcolor=blue]{hyperref}
\usepackage[pdftex]{graphicx}
\usepackage{times,txfonts,ulem}
\usepackage{braket}
\usepackage{color}
\usepackage{natbib}
\usepackage{amsmath,blkarray}
\usepackage{mathtools}
\usepackage{latexsym}
\usepackage{tabularx, booktabs}
\usepackage{graphics,epstopdf}
\usepackage{graphicx}
\usepackage{float}
\usepackage{graphicx}
\usepackage{amsfonts}
\usepackage{subcaption}
\usepackage{color,soul}

\newcommand{\be}{\begin{equation}}
\newcommand{\ee}{\end{equation}}
\newcommand{\ba}{\begin{eqnarray}}
\newcommand{\ea}{\end{eqnarray}}

\begin{document}
\title{Enhanced quantum violation of a non-contextual inequality and witnessing quantum dimension }
\author{ Ritwija Roy }
\email{ritwijaroy0@gmail.com}
\author{ Anindya Biswas }
\email{anindya@nitsikkim.ac.in}

\affiliation{Department of Physics, National Institute of Technology Sikkim, Ravangla, Namchi, Sikkim 737139, India.}
\begin{abstract}
We consider a non-contextual inequality in the sequential measurement scenario and derive the optimal quantum violation of it without assuming the dimension of the system. Since the measurement is dichotomic and the dimension of the quantum system is arbitrary, we formulate the concept of degeneracy-breaking~(DB) measurement depending on how many projectors are being used in the sequential measurement. We demonstrate that by increasing the number of projectors involved in the sequential measurement (thereby making the measurement more degeneracy breaking) the quantum violation of non-contextual inequality can be enhanced and can even reach up to its algebraic maximum. We demonstrate that the optimal quantum violations for different number of projectors serves as a quantum dimension witness. 
\end{abstract}
\maketitle
\section{Introduction}
\label{SecI}
No-go theorems play a key role in quantum foundations by showcasing a radical departure of quantum theory from its classical counterpart. Bell's no-go theorem \cite{belllocal} is arguably the most celebrated one which demonstrates that local realist theories cannot mimic all quantum statistics. However, to demonstrate Bell's theorem, entanglement between two systems are necessary.   Another celebrated no-go theorem, the Kochen-Specker~(KS) theorem \cite{bell,kochen}, demonstrates the inconsistency between quantum theory  and the non-contextual models. Importantly, KS theorem can be demonstrated for a single system without requiring entanglement.  

Various versions of KS theorem were proposed suggesting a variety of ingenious proofs \cite{ker,cabello18, peres1,mermin2,cabello2} in fewer quantum dimensions instead of the original KS proof that requires 117 real vectors in three dimensions.  A state-independent KS proof in three dimension using only 13 vectors was proposed~\cite{yuoh}. Following a different line of argument, Mermin~\cite{mermin2} proposed a state-independent proof in four dimensions by improving the work of Peres~\cite{peres1}. Later, Cabello~\cite{cabellosi} proposed a set of interesting non-contextual inequalities. Some of the contextuality proofs that have been proposed have been experimentally tested as well \cite{michler,hasegawa1,hasegawa2,cabello8,hasegawa3,nature,liu,ams}.

In this paper, we consider a non-contextual inequality~\cite{su} involving four correlations in the sequential scenario. Note that, in prepare-and-measure (or sequential) scenarios, nontrivial certification of quantum advantages generally requires an upper bound on the Hilbert space dimension. Without such a constraint, the problem becomes ill-defined or unconstrained~\cite{chaturvedi2018, VanHimbeeck2017}. However, in this paper we derive the optimal quantum violation without assuming the dimension of the quantum system.  

Furthermore, we introduce the notion of DB measurement in a sequential scenario which leads to enhanced quantum violation of non-contextual inequality. Note that we consider dichotomic observables, but the dimension (\(d > 2\)) of the quantum system is arbitrary, leading to the degeneracy of the eigenvalues. In a sequential measurement scenario, DB measurements can be employed in the first measurement of the sequential measurement setup. In standard degeneracy-preserving (DP) measurement, only two projectors are involved. However, in this paper, any number of projectors from 2 to $d$ have been used to perform the measurement.  Interestingly, we demonstrate that the quantum violation of non-contextual inequality increases with the increase of the number of projectors and can achieve an algebraic maximum for a sufficiently large number of projectors.  Note again that the proof of optimal quantum value of the inequality for a given number of projectors is also presented without putting any upper bound on the dimension $d$ of the quantum system. Additionally, we demonstrate how optimal quantum violations for different number of projectors serve as dimension witness.
\section{A non-contextual inequality and its optimal quantum violation}
\label{secII}

We consider the following algebraic expression 
\begin{eqnarray}\label{l01}
\Delta = A_1A_2  + A_2A_3 +A_3A_4 -A_4A_1 
\end{eqnarray}
where  $A_i \ \forall i \in[4]$ are four dichotomic observables, each having eigenvalues $\pm 1$. Also, they satisfy the following commutation relation $[A_i,A_{i\oplus_4 1}]=0$.

In a realist model, the hidden variable $\lambda$ assigns the value of the observables, so that the measurement outcome  $v(A_i, \lambda)=\pm 1$~\cite{belllocal}. Since $A_1$ commutes with $A_2$, in a non-contextual realist model, $v(A_1, \lambda)$ is independent of $v(A_2, \lambda)$ so that $v(A_1 A_2, \lambda)$ can be written as $v(A_1, \lambda) v(A_2, \lambda)$. Also, for a given  value of $\lambda$, $v(A_1, \lambda)$ remains the same irrespective of the fact that it is paired with $v(A_2, \lambda)$ or $v(A_4, \lambda)$. A similar argument holds for other cases. This gives the non-contextual bound $\langle\Delta\rangle_{nc}\leq 2$.

In Ref.~\cite{su}, by assuming the two-qubit system the maximum quantum value of $\langle\Delta\rangle^{opt}=2\sqrt{2}$ was derived. We first demonstrate that this value is optimal without assuming the dimension of the quantum system. 

For the dimension-independent derivation of the optimal quantum value, we use the sum-of-squares approach~\cite{prabuddha}. We introduce a suitable positive semi-definite operator $\Gamma$ such that the operator $\Delta$ can be written as $\Delta=\beta-\Gamma$. Since $\Gamma\geq 0$, we find that the optimal value of $\Delta$ is obtained when $\Gamma=0$. This, in turn, provides $\Delta^{opt}=\beta$. We define the operator $\Gamma$ as 
\begin{eqnarray}\label{l02}
    \Gamma= \frac{1}{2}(\omega_1 \mathcal{L}_{1}^\dagger\mathcal{L}_{1}+\omega_2 \mathcal{L}_{2}^\dagger\mathcal{L}_{2})
\end{eqnarray}  
where 
\ba
\label{l03}\mathcal{L}_{1}=\frac{A_2-A_{4}}{\omega_{1}} - A_1; \ \ 
\mathcal{L}_{2}=\frac{A_2+A_{4}}{\omega_{2}} - A_3
\ea

The coefficients $\omega_{1}$ and $\omega_2$ are suitable  norms defined as $\omega_{1}=||{A}_{2}-{A}_{4}||_{\rho}=\sqrt{2-\langle\{A_{2},A_{4}\}\rangle}$ and $\omega_{2}=||{A}_{2}+{A}_{4}||_{\rho}=\sqrt{2+\langle\{A_{2},A_{4}\}\rangle}$. Substituting $\mathcal{L}_{1}$ and $\mathcal{L}_{2}$ from Eq.~(\ref{l03}) into Eq.~(\ref{l02}), we get 
\ba \label{l04}\langle\Gamma\rangle=\omega_{1} +\omega_2 -\langle\Delta\rangle
\ea
Therefore, the optimal quantum value of \(\langle\Delta\rangle\) is achieved when  $\langle\Gamma\rangle=0$,  i.e.,
\ba \label{l05}
\langle\Delta\rangle^{opt}&=&\max\left(\omega_{1} +\omega_2\right)\\
&=&
\nonumber
\sqrt{2-\langle\{A_{2},A_{4}\}\rangle}+\sqrt{2+\langle\{A_{2},A_{4}\}\rangle}
\ea
Therefore,  $\langle\Delta\rangle^{opt}=2\sqrt{2}$, when $\{A_{2},A_{4}\}=0$ and $\omega_1=\omega_2=\sqrt{2}$. 

The condition $\langle\Gamma\rangle=0$ also leads to $\langle\psi|\mathcal{L}_{1}^\dagger\mathcal{L}_{1}|\psi\rangle=\langle\psi|\mathcal{L}_{2}^\dagger\mathcal{L}_{2}|\psi\rangle=0$, which further implies, $A_1|\psi\rangle=\frac{A_2-A_{4}}{\sqrt{2}}|\psi\rangle$ and $A_3|\psi\rangle=\frac{A_2+A_{4}}{\sqrt{2}}|\psi\rangle$. Since $A_{2}$ and $A_{4}$ are anti-commuting, it can be checked that $\{A_{1},A_{3}\}=0$.

Note that the derivation does not refer to the dimension of the Hilbert space. However, for setting up this non-contextual inequality, the pairwise commutativity of the observables is necessary. The minimal Hilbert space dimension required to achieve the optimal violation is $d=4$. In 
$d=2$, although dichotomic observables exist, the compatibility structure is too restricted to realize the required set of commuting observables generating the contexts. In 
$d=3$, while richer compatibility structures are available, it is impossible to construct a set of dichotomic observables satisfying the necessary anti-commutation relations. In such a scenario, we derive the specific state which gives the optimal quantum violation. Using the aforementioned conditions on the observables, we derive the quantum state in $d=4$ which can be written as  
\ba
\label{l06}
\rho=\frac{1}{4}\Big[\mathbb{I}_{2}\otimes\mathbb{I}_{2}+\frac{A_1(A_2-A_4)}{\sqrt{2}} +\frac{A_3(A_2+A_4)}{\sqrt{2}} + C_1\Big] 
\ea
where $C_1 = (A_1(A_2-A_4)A_3(A_2+A_4))/2$.

Note that the quantum optimal value cannot be increased further for a higher dimensional system if DP measurement is performed. We  demonstrate that the DB measurement scheme for a higher dimensional system can however enhance the maximum quantum value of \(\langle\Delta\rangle\) and can even reach the algebraic maximum for sufficiently higher dimension. 

\section{Degeneracy breaking measurements}
\label{secIII}
Before proceeding further, let us revisit the essence of degeneracy-preserving (DP) and degeneracy-breaking (DB) measurements, which are encapsulated in the L$\ddot{u}$ders rule and the von Neumann rule, respectively~\cite{luder1}. Consider a dichotomic observable $A$ having  eigenvalues $a\in\{+1,-1\}$ with degree of degeneracy $x_a\in \{1,2....m_a\}$, so that the dimension of the quantum system is $d=\sum_{a}x_{a}$. Let  $A_{a}^{m_{a}}=|\phi_{a}^{m_{a}}\rangle\langle\phi_{a}^{m_a}|$ be the $m_{a}$th rank-1 projector corresponding to the eigenvalue $a$ which gives $A_{a}=\sum\limits_{x_{a}=1}^{m_a}|\phi_{a}^{x_{a}}\rangle\langle\phi_{a}^{x_a}|$. The state reduction may be obtained by considering complete DB measurement employing $A_{a}^{m_{a}}$ or  DP measurement employing $A_{a}$. The post-measurement state for full DB measurements can be written as 
\ba \label{l07} \rho_{db} = \sum_{a,x_a}{A_{a}^{x_a}\rho A_{a}^{x_a}}
\ea
where   $\rho$ is the initial density matrix of the system. On the other hand, in the DP measurement scenario, the post-measurement state can be written as $\rho_{dp}=\sum_{a}{A_{a}\rho A_{a}}$  \cite{heger, pan,kumari18}. One may employ partially DB measurements for which the state reduction can also be derived.  

In this work, for convenience, we consider the dimension of the system $d=2^{n}$ where $n$ is an arbitrary positive integer. We also consider that  for $a=+1$ and $a=-1$ the degree of degeneracy is the same, that is, $m_{+1}=m_{-1}=d/2$. In other words, for both the eigenvalues, the number of rank-1 projectors is the same. 

Let us now assume two dichotomic degenerate observables $\hat{A}$ and $\hat{B}$ in an arbitrary $d$ dimensional system where $A_{a}^{m_a}$ and $B_{b}^{m_{b}}$ are their respective rank-1 projectors. Here $m_a$, $m_b$ are degeneracies corresponding to the eigenvalues $a$ and $b$  of $\hat{A}$ and $\hat{B}$ respectively. The  sequential correlation of $\hat{A}$ and $\hat{B}$ can be written as
\begin{eqnarray}
{\langle \hat{A} \hat{B}\rangle} &=& \sum_{a,b=\pm {1}}a b P({a}{b}),
\label{l08}
\end{eqnarray}
where \(P({a}{b})\) is the probability of getting the eigenvalues \(a\) and \(b\) sequentially.
Using DP measurement the joint probability is given by
\begin{equation} \label{l09}
P_{dp}(a b)= Tr[{{{A_a}}}\rho {A_a}{B_b}]
\end{equation}  where $A_a=\sum_{x_a}A_{a}^{x_{a}}$.  On the other hand, using DB measurement the joint probability is given by 
\begin{equation}
\label{l10}
P_{db}(a b)= Tr\Big[\sum_{x_a=1}^{m_a}{A_{a}^{{x_a}}} \rho A_{a}^{x_a} B_{b}\Big].
\end{equation}
We derive a general relation between the correlations in DB and DP measurements for the case when the dichotomic observable has the same degree of degeneracy  ${\frac{d}{2}}$ corresponding to the 2 eigenvalues. The relation is as follows.
\ba
\label{l11}
&&\langle AB \rangle_{db}=\langle AB\rangle_{dp}
-\\
\nonumber
&&Tr\left[\left(\sum_{x_{+}=1}^{\frac{d}{2}}A_{+}^{x_{+}}\rho\left(\sum_{\substack{x^{\prime}_{+}=1 \\ x_{+}\neq x^{\prime}_{+}}}^{\frac{d}{2}}A_{+}^{x^{\prime}_{+}}\right)-\sum_{x_{-}=1}^{\frac{d}{2}}A_{-}^{x_{-}}\rho\left(\sum_{\substack{x^{\prime}_{-}=1 \\ x_{-}\neq x^{\prime}_{-}}}^{\frac{d}{2}}A_{-}^{x^{\prime}_{-}}\right)\right)B\right]
\ea
We now demonstrate how different values of $d$ lead to different optimal quantum values of $\langle\Delta\rangle$.
\section{Optimal quantum violation for $d=4$}
\label{secIV}
We first derive the optimal value of $\langle\Delta\rangle$ for a quantum system in $d=4$. The observables $A_i$s can be decomposed into four rank-1 projectors, with projectors $A_{i+}^{1}$ and $A_{i+}^{2}$ belonging to the positive eigen-subspace of eigenvalue $+1$, while $A_{i-}^{1}$ and $A_{i-}^{2}$ belong to the negative eigen-subspace with eigenvalue $-1$. From Eq.~(\ref{l11}) the correlation $\langle A_1A_2 \rangle_{db}$ in the  DB measurement can be written as
\begin{align}
    \label{l12}
    \langle A_1A_2 \rangle_{db} = \langle A_1A_2 \rangle_{dp}-
Tr[(A_{1+}^1\rho A_{1+}^2+A_{1+}^2 \rho A_{1+}^1 \\
\nonumber
-A_{1-}^1 \rho A_{1-}^2-A_{1-}^2 \rho A_{1-}^1)A_2]
\end{align}
An extra term arises due to the DB measurement which is in general non-zero and depends on the choice of measurements. We write the rank-1 projectors in the following form.  
\ba \label{l13}
A_{1+}^1=\frac{1}{4}(\mathbb{I}_{4} + A_{1}+M-N), \ \ A_{1+}^2=\frac{1}{4}(\mathbb{I}_4 +A_1-M+N) \\
\nonumber
A_{1-}^{1}=\frac{1}{4}(\mathbb{I}_{4}-A_1+M+N), \ \  A_{1-}^{2}=\frac{1}{4}(\mathbb{I}_{4}-A_1-M-N)
\ea
where $\{A_1,M,N\}$ represents a set of commuting observables with $M,N$ being  suitable dichotomic, Hermitian operators and also satisfying $A_1M=MA_1=N$, $A_1N=NA_1=M$ and $MN=NM=A_1$. Note that the projectors are so designed as to abide by the fundamental rules of operator algebra, given by $\sum_{j}A_{1+}^{j}-\sum_{j}A_{1-}^{j}=A_1$ and $\sum_{j}A_{1+}^{j}+\sum_{j}A_{1-}^{j}=\mathbb{I}_4$. The mutual orthogonality of observables can also be verified easily.

Now, by using Eq.~(\ref{l12}) the correlation  $\langle A_1A_2 \rangle_{db}$ in DB measurement can be derived as   
\begin{equation} \label{l14}
    \langle A_{1} A_{2} \rangle_{db} = \frac{\langle A_1A_2 \rangle_{dp}}{2} - \frac{1}{2}\langle N\alpha N\rangle
\end{equation}
where $\alpha=A_1A_2$ is taken for convenience of notation. 

Similarly, for $\langle A_{2} A_{3} \rangle$, $\langle A_{3} A_{4} \rangle$ and $\langle A_{4} A_{1} \rangle$, we can perform the DB measurements by considering four projectors constructed by the sets of commuting observables $\{A_2, R, S\}$, $\{A_3, E,F\}$ and $\{A_4, U,T\}$ respectively. Therefore, 
\begin{equation}
\label{l15}
\langle\Delta\rangle_{db}^{d=4} = \frac{\langle \Delta\rangle_{dp} }{2}- \frac{\langle N\alpha N+S\beta S+F\gamma F-T\delta T\rangle}{2}
  \end{equation}
where we take $A_2A_3= \beta$, $A_3A_4=\gamma$ and $A_4A_1=\delta$.
Note that the optimal value of $\langle \Delta\rangle_{dp}$ is already derived in Sec.~\ref{secII} which is $\langle \Delta\rangle_{dp}^{opt}=2\sqrt{2}$. To obtain the optimal quantum value in  DB measurement, we need to show that each of the correlations in the second term of Eq.~(\ref{l15}) contributes $\pm1$ suitably for the state in Eq.~(\ref{l06}) and observables $A_{i}$ such that $\langle \Delta\rangle_{db}^{opt}$ can be reached. 

For a two-qubit system we choose $A_1=\sigma_x\otimes \sigma_x$, $A_2=\sigma_z\otimes \sigma_y$,  $A_3=\sigma_x \otimes\sigma_z$ and $A_4=\sigma_y\otimes\sigma_y$ which gives $\alpha=A_1A_2=\sigma_y\otimes \sigma_z$, $\beta=A_2A_3=-\sigma_y\otimes\sigma_x$, $\gamma=A_3A_4=\sigma_z\otimes\sigma_x$ and $\delta=A_4A_1=-\sigma_z\otimes\sigma_z$. Following Eq. (\ref{l06}) the state can be designed as 

\begin{eqnarray}
\label{l16}
\rho_{d=4}=\frac{1}{4}\Big[\mathbb{I}_{2}\otimes\mathbb{I}_{2}+\frac{(\sigma_z+\sigma_y)}{\sqrt{2}}\otimes \sigma_z+\frac{(\sigma_z-\sigma_y)}{\sqrt{2}}\otimes \sigma_x \\
\nonumber
-\sigma_x \otimes\sigma_y \Big]
\end{eqnarray} 

It is straightforward to show that if we choose $N=-cos\frac{\pi}{8}\sigma_z\otimes\sigma_z+sin\frac{\pi}{8}\sigma_y\otimes\sigma_z$ then for the state in Eq.~(\ref{l16}) $\langle N\alpha N\rangle=-1$. Similarly, choosing $S=sin\frac{\pi}{8}\sigma_y\otimes\sigma_x+cos \frac{\pi}{8}\sigma_z\otimes\sigma_x$, $F=sin \frac{\pi}{8}\sigma_z\otimes\sigma_x+cos \frac{\pi}{8}\sigma_y\otimes\sigma_x$ and   $T = cos\frac{\pi}{8}\sigma_y\otimes\sigma_z-sin \frac{\pi}{8}\sigma_z\otimes\sigma_z$ we get $\langle S\beta S\rangle=\langle F\gamma F\rangle=-\langle T\delta T\rangle=-1$. Hence, the maximum value that can be attained with the DB measurement scheme is
 \ba
(\langle\Delta\rangle^{d=4}_{db})^{opt}=\sqrt{2}+2
 \ea
 which is higher than $2\sqrt{2}$ and establishes the supremacy of the DB measurement over the DP measurement in this context.
 \section{Optimal quantum violation for $d=8$}
\label{secV}
Now, we consider the full DB measurement for the system with dimension $d=8$. We also take that each of the eigen subspaces is four-fold degenerate. Following Eq.~(\ref{l11}) the sequential correlation in the DB measurement gives rise to the relation
\begin{equation} \label{l18}
    \begin{split}
    \langle A_1A_2 \rangle_{db}=\langle A_1A_2 \rangle_{dp}
    -Tr[(A_{1+}^{1}\rho(A_{1+}^{2}+A_{1+}^{3}+A_{1+}^{4})\\
    +A_{1+}^{2}\rho(A_{1+}^{1}+A_{1+}^{3}+A_{1+}^{4}) \\
+A_{1+}^{3}\rho(A_{1+}^{1}+A_{1+}^{2}+A_{1+}^{4})\\
+A_{1+}^{4}\rho(A_{1+}^{1}+A_{1+}^{2}+A_{1+}^{3})\\
-A_{1-}^{1}\rho(A_{1-}^{2}+A_{1-}^{3}+A_{1-}^{4})\\
-A_{1-}^{2}\rho(A_{1-}^{1}+A_{1-}^{3}+A_{1-}^{4})\\
-A_{1-}^{3}\rho(A_{1-}^{1}+A_{1-}^{2}+A_{1-}^{4})\\
-A_{1-}^{4}\rho(A_{1-}^{1}+A_{1-}^{2}+A_{1-}^{3}))A_{2}]
         \end{split}
\end{equation}
where $A_{1+}^{j}$ and $A_{1-}^{j}$s are rank-1 projectors corresponding to eigenvalues $+1$ and $-1$ respectively $\forall j \in [4]$. The projectors can be written as
 
\ba
\label{l19}
A_{1+}^1=\frac{1}{8}[\mathbb{I}_{8}+A_1+N_1+N_2+N_3+N_4+N_5+N_6] \\
\nonumber
A_{1+}^2=\frac{1}{8}[\mathbb{I}_{8}+A_1-N_1-N_2+N_3+N_4-N_5-N_6] \\
\nonumber
A_{1+}^3=\frac{1}{8}[\mathbb{I}_{8}+A_1-N_1-N_2-N_3-N_4+N_5+N_6] \\
\nonumber
A_{1+}^4=\frac{1}{8}[\mathbb{I}_{8}+A_1+N_1+N_2-N_3-N_4-N_5-N_6] \\
\nonumber
A_{1-}^1=\frac{1}{8}[\mathbb{I}_{8}-A_1-N_1+N_2-N_3+N_4-N_5+N_6] \\
\nonumber
A_{1-}^2=\frac{1}{8}[\mathbb{I}_{8}-A_1+N_1-N_2-N_3+N_4+N_5-N_6] \\
\nonumber
A_{1-}^3=\frac{1}{8}[\mathbb{I}_{8}-A_1+N_1-N_2+N_3-N_4-N_5+N_6] \\
\nonumber
A_{1-}^4=\frac{1}{8}[\mathbb{I}_{8}-A_1-N_1+N_2+N_3-N_4+N_5-N_6]
\ea
where $N_i$, $\forall i \in[6]$ are mutually commuting dichotomic Hermitian observables. 

Here also we can validate the projectors with the help of spectral-decomposition ($\sum_{j}A_{1+}^{j}-\sum_{j}A_{1-}^{j}=A_1$) and completeness theorem ($\sum_{j}A_{1+}^{j}+\sum_{j}A_{1-}^{j}=\mathbb{I}_8$) together, while they are mutually orthogonal as well. This fixes  that $A_1N_1=N_1A_1=N_2$ , $A_1N_3=N_3A_1=N_4$ , $A_1N_5=N_5A_1=N_6$. 

Putting Eq.~(\ref{l19}) into Eq.~(\ref{l18}), we derive 
\ba
\label{l20}
\langle A_1 A_2 \rangle _{db}= \frac{\langle A_1A_2 \rangle_{dp}}{4}+\frac{1}{4}\langle N_2 \alpha N_2+N_4 \alpha N_4+N_6 \alpha N_6\rangle
\ea
 As we have done for $d=4$, to obtain the maximum quantum violation we need to show that $\langle N_2 \alpha N_2\rangle=\langle  N_4 \alpha N_4\rangle=\langle N_6 \alpha N_6\rangle=1$  for a state $\rho$ with $\{A_1, N_2, N_4, N_6\}$ being mutually commuting. The state can be derived as 

\begin{equation}
\label{l21}
    \begin{split}
        \rho=\frac{1}{8}\Big[\mathbb{I}_8+\frac{A_1(A_2-A_4)}{\sqrt{2}}+\frac{A_3(A_2+A_4)}{\sqrt{2}}+\sum_{i=1}^{5}C_{i}\Big]
    \end{split}
\end{equation}
Here $\{C_{i}\}$ is the set of mutually commuting observables which also commute with $\frac{A_1(A_2-A_4)}{\sqrt{2}}$ and $\frac{A_3(A_2+A_4)}{\sqrt{2}}$.

We give an explicit example of states and observables in the three-qubit system that leads to the optimal value of \(\langle\Delta\rangle\).  Let, $A_1=\sigma_x\otimes\sigma_x\otimes\sigma_x$, $A_2=\sigma_z\otimes\sigma_x\otimes\sigma_z$, $A_3=\sigma_y\otimes\sigma_x\otimes\sigma_x$ and $A_4=\sigma_z\otimes\sigma_x\otimes\sigma_y$. The operators thus defined satisfy the requirements of all commutativity as well as of anti-commutativity. We can further write $A_1A_2=\alpha=-\sigma_y\otimes \mathbb{I}\otimes\sigma_y$, $A_2A_3=\beta=\sigma_x\otimes\mathbb{I}\otimes\sigma_y$, $A_3A_4=\gamma=-\sigma_x\otimes\mathbb{I}\otimes\sigma_z$ and $A_4A_1=\delta=\sigma_y\otimes\mathbb{I}\otimes\sigma_z$. Then the state given by Eq.~\ref{l21} can explicitly be written as 
\begin{eqnarray}
\label{l22}
\nonumber
 \rho_{d=8}&&=\frac{1}{8}\big[\mathbb{I}_8-\frac{\sigma_y\otimes\mathbb{I}\otimes(\sigma_y+\sigma_z)}{\sqrt{2}}+\frac{\sigma_x\otimes\mathbb{I}\otimes(\sigma_y-\sigma_z)}{\sqrt{2}}\\
&&-\frac{\sigma_y\otimes\sigma_x\otimes(\sigma_y+\sigma_z)}{\sqrt{2}}+\frac{\sigma_x\otimes\sigma_x\otimes(\sigma_y-\sigma_z)}{\sqrt{2}}\\
\nonumber
&&+\sigma_z\otimes\mathbb{I}\otimes\sigma_x+\mathbb{I}\otimes\sigma_x\otimes\mathbb{I}+\sigma_z\otimes\sigma_x\otimes\sigma_x\big]   
\end{eqnarray}
Now, we choose the operators  $N_2=sin\frac{\pi}{8}\sigma_x\otimes\sigma_z\otimes\sigma_y - cos\frac{\pi}{8}\sigma_x\otimes\sigma_z\otimes\sigma_z $, $N_4 = sin \frac{\pi}{8}\mathbb{I}_2\otimes\sigma_y\otimes\sigma_z + cos \frac{\pi}{8}\mathbb{I}_2\otimes\sigma_y\otimes\sigma_y $ and $N_6=cos\frac{\pi}{8}\mathbb{I}_2\otimes\sigma_z\otimes\sigma_y + sin\frac{\pi}{8}\mathbb{I}_2\otimes\sigma_z\otimes\sigma_z $. For the state in Eq.~(\ref{l22}) we get
\ba
\langle N_2 \alpha N_2\rangle= \langle N_4 \alpha N_4\rangle= \langle N_6 \alpha N_6\rangle=1
\ea
Thus we finally get
\begin{eqnarray}
\langle A_1A_2 \rangle_{db} = \frac{\langle A_1A_2 \rangle_{dp}}{4}+\frac{3}{4}   
\end{eqnarray}
Deriving the other three correlations in a similar way with the choice of proper observables, we find

\begin{equation}
\label{dbfinal1}
\langle \Delta \rangle^{d=8}_{db} = \frac{\langle \Delta\rangle_{dp} }{4}+3
  \end{equation}
Since $(\langle \Delta\rangle_{dp})^{opt}=2\sqrt{2}$, we finally have 
\ba
\label{eq:23}
(\langle \Delta \rangle^{d=8}_{db})^{opt} = \frac{1}{\sqrt{2}}+3 \approx 3.707
\ea
which is again higher than $(\langle \Delta \rangle^{d=4}_{db})^{opt}$.

\section{Generalization for arbitrary $d=2^{n}$ }
\label{VI}

Now, we extend our framework to higher dimensional systems by breaking the degeneracy even further. Following the same techniuqe as shown earlier we have checked for $d=16$, $(\langle \Delta \rangle _{db}^{d=16})^{opt}=\frac{1}{2\sqrt{2}}+\frac{7}{2}\approx 3.854$. Thus the optimal violation for an arbitrary system of dimension $d=2^{n}$ with the proper choice of observables can be found by mathematical induction to be 

\ba
\label{l27}
(\langle \Delta \rangle _{db}^{d=2^{n}})^{opt}=\frac{\langle \Delta \rangle _{dp}^{opt}}{2^{n-1}}+\frac{(2^{n-1}-1)}{2^{n-3}}
\ea
The required state can be derived as 
\begin{equation}\label{l28}
    \begin{split}
        \rho_{d=2^{n}}=\frac{1}{2^n}\Big[\mathbb{I}_{2^n}+\frac{A_1(A_2-A_4)}{\sqrt{2}}+\frac{A_3(A_2+A_4)}{\sqrt{2}}\\+\sum_{k=1}^{2^{n}-3} \eta_{k}C_{k}\Big]
    \end{split}
\end{equation}
Here, all the components of $\rho_{d=2^n}$ must follow idempotency and should mutually commute, also $\eta_{k} \in \{\pm1\}$. $C_{k}$'s can be found by properly multiplying the preceding terms, satisfying commutativity. From Eq.~(\ref{l27}) it is evident that for a large value of $n$, the first term goes to zero and we get $(\langle \Delta \rangle _{db}^{d=2^{n}})^{opt} \approx 4$,  which is the algebraic maximum of $\Delta$. Therefore, we demonstrate extreme quantum violation of the non-contextualiy inequality by employing the DB measurement  scheme.
\section{Witnessing quantum dimension}
\label{secVII}
The dimension of the system is a fundamental property and is considered to
be a resource for quantum information theory where a higher dimensional system may produce a higher degree of non-classicality. The quantum dimension witness fixes a lower bound on the dimension that is needed to reproduce a given measurement statistics. Over the years, many forms of dimension witnesses have been formulated  \cite{brunner,gallego,guhne,bowles,sikora,ahrens,brunner1,Wolf,pan1}. We demonstrate here how the optimal quantum violation for a given number of projectors serves as a dimension witness.

\begin{table}[htbp]
\centering
\caption{Optimal value of $\langle \Delta \rangle_{db}^{opt}$ for varying dimensions}
\label{tab1}
\renewcommand{\arraystretch}{1.8} 
\setlength{\tabcolsep}{6pt}

\begin{tabular}{|l|c|c|c|c|}
\hline
Dimension & {$d=4$} &~ {$d=8$} &~ {$d=16$} & ~{$d\rightarrow\infty$} \\
\hline
$\langle \Delta \rangle_{db}^{opt}$ & 3.414 &~ 3.707 &~ 3.854 &~ 4 \\
\hline
\end{tabular}
\end{table}



 The results are summarized in Table~(\ref{tab1}). For example, $(\langle \Delta \rangle _{db}^{d=8})^{opt}=3.707 $ can only be achieved if the dimension of the Hilbert space is $d\geq 8$. This is due to the fact that one needs at least eight projectors to perform the DB measurement to achieve $(\langle \Delta \rangle _{db}^{d=8})^{opt}$, and hence the measurement result fixes the minimum dimension to be eight.  If the dimension is less than eight, then this bound cannot be achieved. The same holds for other optimal values.
\section{Summary and discussions}
\label{secVII}
We derived the optimal quantum violation of a non-contextual inequality in sequential measurement scenario. In contrast to the conventional belief, we demonstrated a dimension independent derivation of the optimal quantum value by using an elegant sum-of-square approach. This fixes the relation between the observables and the state required for the optimal violation. 

Using DB measurement, we showed how the optimal quantum value obtained for DP measurement can be enhanced.  We specifically demonstrated that higher the number of projectors for the  DB measurement, higher the optimal quantum violation that can be achieved, and for a sufficiently large number of projectors the algebraic maximum of the functional $\langle\Delta\rangle$ can be reached. Furthermore, we showed how different optimal quantum violations for different numbers of projectors can serve as dimension witnesses. Finally, we note that the methodology outlined here can be used for other forms of inequality for achieving enhanced quantum inequality. Study along such a line could be an interesting avenue for future research. 

\section*{Acknowledgements}
The authors thank A. K. Pan for proposing the problem and insightful, rigorous discussions that helped to enrich the paper extensively.


\begin{thebibliography}{99}

\bibitem{belllocal}J. S. Bell, \textit{``On the Einstein Podolsky Rosen paradox'' }, \href{https://doi.org/10.1103/PhysicsPhysiqueFizika.1.195 }{Physics, \textbf{1} 195 (1964)}.
\bibitem{bell}J. S. Bell, \textit{``On the Problem of Hidden Variables in Quantum Mechanics''}, \href{ https://doi.org/10.1103/RevModPhys.38.447}{ Rev. Mod. Phys. \textbf{38}, 447 (1966)}.
\bibitem{luder1}G. Lüders, \textit{``Über die Zustandsänderung durch den Meßprozeß''}, \href{https://doi.org/10.1002/andp.19504430510}{Annalen der Physik, {\bf443}, (1950)}.
\bibitem{kochen}S. Kochen, and E. P. Specker, \textit{``The Problem of Hidden Variables in Quantum Mechanics''}, \href{https://www.jstor.org/stable/24902153 }{ J. Math. Mech. \textbf{17}, 59 (1967)}.

\bibitem{ker}M. Kernaghan, and A. Peres, \textit{``Kochen-Specker theorem for eight-dimensional space''}, \href{https://doi.org/10.1016/0375-9601(95)00012-R}{ Phys. Lett. A \textbf{198}. 1 (1995)}.
\bibitem{cabello18}A. Cabello, J. M. Estebaranz, and G. Garcia-Alcaine, \textit{``Bell-Kochen-Specker theorem: A proof with 18 vectors''}, \href{
https://doi.org/10.1016/0375-9601%2896%2900134-X}{ Phys. Lett. A \textbf{212}, 183 (1996)}.

\bibitem{peres1}A. Peres, \textit{``Two simple proofs of the Kochen-Specker theorem''}, \href{https://doi.org/10.1088/0305-4470/24/4/003}{J. Phys. A \textbf{24}, L175 (1991)}.

\bibitem{mermin2}N. D. Mermin, \textit{``Hidden variables and the two theorems of John Bell''}, \href{https://doi.org/10.1103/RevModPhys.65.803}{Rev. Mod. Phys. \textbf{65}, 803 (1993)}.

\bibitem{cabello2}A. Cabello, and G. Garcia-Alcaine, \textit{``Proposed Experimental Tests of the Bell-Kochen-Specker Theorem''}, \href{https://doi.org/10.1103/PhysRevLett.80.1797}{Phys. Rev. Lett. \textbf{80}, 1797 (1998)} 
\bibitem{yuoh} S. Yu and C. H. Oh, \textit{``State-Independent Proof of Kochen-Specker Theorem with 13 Rays''}, \href{ https://doi.org/10.1103/PhysRevLett.108.030402}{Phys. Rev. Lett. {\bf108}, 030402 (2012)} 
\bibitem{cabellosi} A. Cabello, \textit{``Experimentally Testable State-Independent Quantum Contextuality''}, \href{ https://doi.org/10.1103/PhysRevLett.101.210401}{Phys. Rev. Lett. {\bf 101} 210401 (2008)}.

\bibitem{michler}M. Michler, H. Wienfurter and M. Zukowski, \textit{``Experiments towards Falsification of Noncontextual Hidden Variable Theories''}, \href{https://doi.org/10.1103/PhysRevLett.84.5457 }{PHys.Rev.Lett. \textbf{84} 5457 (2000)}.
\bibitem{hasegawa1}Y. Hasegawa, R. Loidl, G. Badurek, M. Baron, and H. Rauch, \textit{``Violation of a Bell-like inequality in single-neutron interferometry''}, \href{https://doi.org/10.1038/nature01881}{Nature \textbf{425}, 45 (2003)}. J. Opt. B: Quantum Semiclass. Opt. \textbf{6}, S7-S12 (2004).
\bibitem{hasegawa2}Y. Hasegawa, R. Loid, G. Badurek, M. Baron, and H. Rauch, \textit{``Quantum Contextuality in a Single-Neutron Optical Experiment''}, \href{  https://doi.org/10.1103/PhysRevLett.97.230401}{Phys. Rev. Lett. \textbf{97}, 230401 (2006)}. 
\bibitem{cabello8}A. Cabello, S. Filipp, H. Rauch, and Y. Hasegawa, \textit{``Proposed Experiment for Testing Quantum Contextuality with Neutrons''}, \href{ https://doi.org/10.1103/PhysRevLett.100.130404}{Phys. Rev. Lett. \textbf{100}, 130404 (2008)}.
\bibitem{hasegawa3} H. Bartosik, J. Klepp, C. Schmitzer, S. Sponar, A. Cabello, H. Rauch1,, and Y. Hasegawa1, \textit{``Experimental Test of Quantum Contextuality in Neutron Interferometry''}, \href{ https://doi.org/10.1103/PhysRevLett.103.040403}{Phys. Rev. Lett. {\bf 103} 040403(2009)}.  

\bibitem{nature} G. Kirchmair, F. Zähringer, R. Gerritsma, M. Kleinmann, O. Gühne, A. Cabello, R. Blatt and C. F. Roos, \textit{``State-independent experimental test of quantum contextuality''}, \href{ https://doi.org/10.1038/nature08172}{Nature {\bf 460} 494 (2009)} .
\bibitem{liu} B. H. Liu, Y. F. Huang, Y. X. Gong, F. W. Sun, Y. S. Zhang, C. F. Li, and G. C. Guo, \textit{``Experimental demonstration of quantum contextuality with nonentangled photons''}, \href{ https://doi.org/10.1103/PhysRevA.80.044101}{ Phys. Rev. A, {\bf 80} 044101(2009)}.
\bibitem{ams}E. Amselem, M. Radmark, M. Bourennane and A. Cabello, \textit{``State-Independent Quantum Contextuality with Single Photons''}, \href{https://doi.org/10.1103/PhysRevLett.103.160405 }{ Phys. Rev. Lett. {\bf 103}, 160405 (2009)}.
\bibitem{su}H-Y. Su, J-Lg Chen, C. Wu, S. Yu, and C. H. Oh, \textit{``Quantum contextuality in a one-dimensional quantum harmonic oscillator''}, \href{ https://doi.org/10.1103/PhysRevA.85.052126}{ Phys. Rev. A {\bf85}, 052126 (2012)}.
\bibitem{chaturvedi2018}A. Chaturvedi, M. Ray, R. Veynar, Ryszard and M. Paw{\l}owski, \textit{``On the security of semi-device-independent QKD protocols''}, \href{https://doi.org/10.1007/s11128-018-1892-z}{Quantum Inf. Process. {\bf17}, 131 (2018)}.
\bibitem{VanHimbeeck2017}T. V. Himbeeck, E. Woodhead, N. J. Cerf,  R. Garc{\'{i}}a-Patr{\'{o}}n, S. Pironio, \textit{``Semi-device-independent framework based on natural physical assumptions''}, \href{https://doi.org/10.22331/q-2017-11-18-33}{Quantum, {\bf1}, (2017)}.
\bibitem{prabuddha} P. Roy and A. K. Pan, \textit{``Device-independent self-testing of unsharp measurements''}, \href{https://doi.org/10.1088/1367-2630/acb4b5}{New Journal of Physics {\bf25}, 013040 (2023)}.
   \bibitem{heger} G. C. Hegerfeldt and R. Sala Mayato, \textit{``Discrimination of measurement contexts in quantum mechanics''}, \href{https://doi.org/10.1016/j.physleta.2011.07.012}{Phy Lett. A \textbf{375}, 3167 (2011)}.

\bibitem{pan} A. K. Pan and K. Mandal, \textit{``Quantum Contextuality for a Three-Level System Sans Realist Model''}, \href{ https://doi.org/10.1007/s10773-016-2974-2}{Int. J. Theor. Phys. \textbf{55}, 3472 (2016)}.
\bibitem{kumari18} A. Kumari, Md. Qutubuddin and A. K. Pan, \textit{``Violation of the Lüders bound of macrorealist and noncontextual inequalities''}, \href{https://doi.org/10.1103/PhysRevA.98.042135}{Phys. Rev. A \textbf{98}, 042135 (2018)}.

\bibitem{brunner}N. Brunner, S. Pironio, A. Acin, N. Gisin, A.A. Méthot, and V. Scarani, \textit{``Testing the Dimension of Hilbert Spaces''},\href{https://doi.org/10.1103/PhysRevLett.100.210503}{Phys. Rev. Lett. {\bf100}, 210503 (2008) }.

\bibitem{Wolf}M. M. Wolf and D. Perez-Garcia, \textit{``Assessing Quantum Dimensionality from Observable Dynamics''},\href{https://doi.org/10.1103/PhysRevLett.102.190504}{Phys. Rev. Lett. {\bf102}, 190504 (2009)}.
\bibitem{gallego}R. Gallego, N. Brunner, C. Hadley, and A. Acín, \textit{``Device-Independent Tests of Classical and Quantum Dimensions''}, \href{ https://doi.org/10.1103/PhysRevLett.105.230501}{Phys. Rev. Lett. {\bf105}, 230501 (2010) }.

\bibitem{ahrens}J. Ahrens, P. Badzi\c{a}g, A. Cabello and Md. Bourennane, \textit{``Experimental device-independent tests of classical and quantum dimensions''}, \href{https://doi.org/10.1038/nphys2333}{Nature Phys {\bf8}, 592–595 (2012)}.


\bibitem{brunner1}N. Brunner, M. Navascués, and T. Vértesi, \textit{``Dimension Witnesses and Quantum State Discrimination''}, \href{https://doi.org/10.1103/PhysRevLett.110.150501}{Phys. Rev. Lett. {\bf110}, 150501 (2013)}.



\bibitem{guhne}O. Gühne, C. Budroni, A. Cabello, M. Kleinmann, and J.-Å. Larsson, \textit{``Bounding the quantum dimension with contextuality''}, \href{https://doi.org/10.1103/PhysRevA.89.062107}{Phys. Rev. A {\bf89}, 062107 (2014) }.
\bibitem{bowles}J. Bowles, M.-T. Quintino, and N. Brunner, \textit{``Certifying the Dimension of Classical and Quantum Systems in a Prepare-and-Measure Scenario with Independent Devices''}, \href{https://doi.org/10.1103/PhysRevLett.112.140407}{Phys. Rev. Lett. {\bf112}, 140407(2014)}.

\bibitem{sikora}J. Sikora, A. Varvitsiotis, and Z. Wei, \textit{``Device-independent dimension tests in the prepare-and-measure scenario''}, \href{https://doi.org/10.1103/PhysRevA.94.042125}{Phys. Rev. A {\bf94}, 042125 (2016) }.





\bibitem{pan1}A. K. Pan and S. S. Mahato, \textit{``Device-independent certification of the Hilbert-space dimension using a family of Bell expressions''}, \href{https://doi.org/10.1103/PhysRevA.102.052221}{Phys. Rev. A {\bf102}, 052221 (2020)}.










\end{thebibliography}
\end{document}